\documentclass[lettersize,journal]{IEEEtran}

\usepackage[noadjust]{cite}
\usepackage{amsmath,amsfonts}
\usepackage{algorithmic}
\usepackage{array}
\usepackage[caption=false,font=normalsize,labelfont=sf,textfont=sf]{subfig}
\usepackage{textcomp}
\usepackage{stfloats}
\usepackage{url}
\usepackage{verbatim}
\usepackage{graphicx}
\usepackage{hyperref}
\usepackage[nameinlink]{cleveref}
\usepackage{amsthm}
\usepackage{xcolor}

\newtheorem{theorem}{Theorem}

\theoremstyle{definition}
\newtheorem{definition}{Definition}
\newtheorem{problem}{Problem}
\newtheorem{remark}{Remark}

\def\equationautorefname~#1\null{(#1)\null}
\def\sectionautorefname~#1\null{Section #1\null}
\def\subsectionautorefname~#1\null{Section #1\null}
\def\theoremautorefname~#1\null{Theorem #1\null}
\def\remarkautorefname~#1\null{Remark #1\null}
\def\problemautorefname~#1\null{Problem #1\null}
\def\definitionautorefname~#1\null{Definition #1\null}

\usepackage{balance}
\begin{document}

\title{Invariance Guarantees using\\Continuously Parametrized Control Barrier Functions}

\author{Inkyu Jang and H. Jin Kim
\thanks{The authors are with the Department of Aerospace Engineering and Automation and Systems Research Institute (ASRI), Seoul National University, 08826 Seoul, Korea (e-mail:{janginkyu.larr@gmail.com; hjinkim@snu.ac.kr}).}%
}

\maketitle

\begin{abstract}
Constructing a control invariant set with an appropriate shape that fits within a given state constraint is a fundamental problem in safety-critical control but is known to be difficult, especially for large or complex spaces. 
This paper introduces a safe control framework of utilizing PCBF: continuously parametrized control barrier functions (CBFs). In PCBF, each choice of parameter corresponds to a control invariant set of relatively simple shape. Invariance-preserving control is done by dynamically selecting a parameter whose corresponding invariant set lies within the safety bound.
This eliminates the need for synthesizing a single complex CBF that matches the entire free space.
It also enables easier adaptation to diverse environments. 
By assigning a differentiable dynamics on the parameter space, we derive a lightweight feedback controller based on quadratic programming (QP), namely PCBF-QP.
We also discuss on how to build a valid PCBF for a class of systems and how to constrain the parameter so that the invariant set does not exceed the safety bound.
The concept is also extended to cover continuously parametrized high-order CBFs, which is called high-order PCBF. 
Finally, simulation experiments are conducted to validate the proposed approach.
\end{abstract}

\section{Introduction}

Ensuring safety when designing a control law for a controlled system is very important in many real-world applications.
In order to address not myopic but persistent satisfaction of the safety requirements, safety should be addressed from the perspective of set invariance \cite{blanchini1999set}. A typical safety-critical control methodology therefore aims for creating a control invariant set that is entirely contained within the pre-given set of allowable states. An invariance-preserving control is then applied to keep the system's state within this set.

One of the most widely used approaches to construct a control invariant set is to utilize a control barrier function (CBF) \cite{ames2019control}. CBF is a Lyapunov-like scalar function defined on the state space, whose super zero level set defines the control invariant set. Its main strength comes from its simplicity of encoding invariance using a single scalar function. In addition, once synthesized, a valid CBF offers a computationally efficient means of enforcing safety constraints through quadratic programming (QP), namely CBF-QP \cite{ames2016control}. CBF-QP introduces one additional inequality constraint to the input bounds and can be solved
in real-time by any off-the-shelf convex programming solver.

These advantages offered by CBF and CBF-QP have drawn researchers' significant attention, leading to an extensive body of literature regarding practical applications, especially in robotics \cite{grandia2021multi, wang2017safety, singletary2022onboard, jang2024safe, oh2023safety} and also in other topics \cite{ames2020safety, feng2023fast}.
There also have been works to employ CBFs in a wider range of applications, for example, safety-critical reinforcement learning \cite{choi2020reinforcement}, control of systems with stochasticity \cite{clark2019control, santoyo2021barrier}, adapting to changing dynamics \cite{xiao2021adaptive}, time-varying CBFs for satisfaction of signal temporal logic specifications \cite{lindemann2018control}.
The recent concept of high-order CBF (HOCBF) \cite{xiao2021high} generalizes the notion of CBF to cover high relative degree.

The common aim of these methods is to construct a barrier certificate corresponding to a invariant set that fits within the given state constraints.
Considering the control performance, it is obviously important to obtain a valid barrier function that provides a control invariant set with appropriate size and shape while not exceeding the prescribed limit.
Unfortunately, finding a control invariant set (or a valid CBF) subject to the state constraints is often not straightforward and easily becomes computationally burdensome, particularly with large or complex environments.

In this paper, we introduce an invariance-preserving control framework that uses continuously parametrized spectrum of CBFs, which we call parametrized CBF (PCBF).
PCBF is a Lyapunov-like scalar function that takes not only the state but also a parameter which lives in a continuous parameter space. For each fixed parameter, PCBF is a valid yet relatively simple CBF that defines a small control invariant set.
Given a set of \textit{safe} parameters (whose corresponding invariant sets lie within the safety bound), a larger and more complex-shaped control invariant set can be constructed by taking the union of the corresponding invariant sets.

The proposed PCBF framework decouples the problem of finding a safe invariant set
into the following two:
\begin{problem} \label{prob:pcbf_design}
    Given the system dynamics, find a continuous spectrum of small, simple-shaped (hence more manageable) control invariant sets, i.e., the building-block invariant sets.
\end{problem}
\begin{problem} \label{prob:parameter_design}
    Given the environmental information, determine which ones among the building-block invariant sets lie within the safety bound, and construct the parameter constraint that describes the safe parameter set.
\end{problem}
\noindent \autoref{prob:pcbf_design} corresponds to finding a valid PCBF; this step does not require considering the potentially complex safety bound.
Given the building-block invariant sets, \autoref{prob:parameter_design} is purely geometric (it is a problem of determining set inclusion) and does not explicitly depend on the system dynamics.
In many cases, this significantly reduces the computational burden compared to directly searching for a single valid barrier certificate that spans the entire free space. Moreover, it gives adaptability to various environments, since only the parameter constraint (and not the PCBF) needs to be resynthesized when shifting to a new safety bound.

Finding a general answer to the two problems (i.e., a systematic way of constructing a PCBF and the parameter constraint from the dynamics model only) is hard and remains an open problem.
However, in this paper, we find that continuous symmetry of the system dynamics, which is often found in a wide range of real-world dynamical systems, naturally extends a single HOCBF to give a valid PCBF. We also introduce a design process of a PCBF using a stabilizing controller and a Lyapunov function. We also briefly discuss on how the parameter constraints should be constructed.

Once a valid PCBF and the parameter constraints are obtained, we are then interested in the invariance-preserving control problem:
\begin{problem} \label{prob:control}
    Synthesize a computationally lightweight feedback control that renders the resulting safe set (the union of the building blocks found in \autoref{prob:parameter_design}) invariant.
\end{problem}
\noindent To address \autoref{prob:control}, we firstly assign a single integrator dynamics on the parameter space and augment it to the system. 
The control is done with respect to the augmented system in a manner that the parameter always stays within the safe parameter set, and the state stays within the building-block invariant set corresponding to the current parameter value.
Leveraging the continuity of the parameter space, we devise a CBF-QP-like safety filter based on PCBF (PCBF-QP), which is capable of constraining the augmented system as mentioned.

This paper generalizes our recent conference paper \cite{jang2023invariance_2} to high-relative-degree outputs. Following the motivation given by HOCBFs \cite{xiao2021high}, high-order PCBF (HOPCBF) is defined as a differentiably parametrized spectrum of HOCBFs, and we derive the set of input constraints for invariance guarantees using a HOPCBF. 
A QP-based safety filter using HOPCBF (HOPCBF-QP) is then proposed, which is capable of generating inputs satisfying the HOPCBF constraints at a low computational cost.

We also discuss the design process of the building-block invariant sets, i.e., \autoref{prob:pcbf_design} and \autoref{prob:parameter_design}, which are new in this journal version. We present numerical examples, the first one demonstrates the barrier function design process abovementioned, the second addresses the high-order case, and the third showcases the applicability of the method to infinitely many safety constraints.

The remainder of this paper is organized as follows.
We start by introducing the necessary concepts and assumptions in \autoref{sec:preliminaries}. The concept of PCBF and its definition is presented in \autoref{sec:pcbf} in which we also derive PCBF-QP and answer \autoref{prob:control}. The concept of PCBF is generalized to high-order in \autoref{sec:hopcbf}. In \autoref{sec:remarks}, we discuss the design process of a (HO)PCBF and give partial answers to \autoref{prob:pcbf_design} and \autoref{prob:parameter_design}. The numerical examples are presented in \autoref{sec:case_study}, which is followed by the summary and outlook of the work in \autoref{sec:conclusion}.

\section{Preliminaries} \label{sec:preliminaries}

\subsection{Notation}

For positive integers $l$, $m$, and $n$, $\mathbb{R}^l$ and $\mathbb{R}^{m\times n}$ denote the set of $l$-dimensional real column vectors and matrices of size $m \times n$, respectively.
An inequality between two vectors denotes that it is satisfied in an elementwise manner.
We use the notation $\partial_{\xi} \beta(\xi)$ to denote the partial derivative of $\beta$ with respect to argument $\xi$.
The Lie derivative of a function $V:\mathbb{R}^n \rightarrow \mathbb{R}$ along a vector field $f:\mathbb{R}^n\rightarrow \mathbb{R}^n$ is written as $L_f V(x) = \partial_x V(x) \cdot f(x)$.
If a function $f$ is $r$ times continuously differentiable, we write $f \in \mathcal{C}^r$.
Throughout this paper, we use the symbols $x\in \mathbb{R}^n$, $u\in \mathbb{R}^m$, $k\in \mathbb{R}^{n_k}$, and $t\in [0,\infty)$ to denote state, input, CBF parameter, and time, respectively. The roman-font $\mathrm{x}$, $\mathrm{u}$, $\mathrm{k}$ are used to emphasize that they are \textit{trajectories}, i.e., functions of time.

\subsection{Dynamics}

In this paper, we consider the following nonlinear time-invariant control-affine system dynamics:
\begin{equation} \label{eq:dynamics}
    \dot{\mathrm{x}}(t) = f(\mathrm{x}(t)) + g(\mathrm{x}(t)) \mathrm{u}(t).
\end{equation}
The functions $f:\mathbb{R}^n\rightarrow \mathbb{R}^n$ and $g:\mathbb{R}^n \rightarrow \mathbb{R}^{n\times m}$ are locally Lipschitz functions so that given any initial value $\mathrm{x}(0)$ and a measurable input trajectory $\mathrm{u}(t)$, there exists a unique solution $\mathrm{x}$ at least locally.
The input is assumed to be bounded by a set of linear inequalities, i.e.,
\begin{equation} \label{eq:input_set_assumption}
    \mathrm{u}(t) \in U = \{u \in \mathbb{R}^m : A_uu\leq b_u\}, \quad \forall t \in [0,\infty)
\end{equation}
where $A_u$ and $b_u$ are a matrix and a column vector with appropriate sizes.

\subsection{Set Invariance and Control Barrier Functions}
Set invariance is a key concept in ensuring safety of a system. We first begin with the definition of control invariant sets.

\begin{definition}[Control Invariant Set]
A set $C \subseteq \mathbb{R}^n$ is control invariant for the system \autoref{eq:dynamics} if for every $\mathrm{x}(0)\in C$, there exists an input trajectory $\mathrm{u}:[0,\infty) \rightarrow U$ that makes the resulting state trajectory reside permanently in $C$, i.e., $\mathrm{x}(t) \in C$, for all $t \in [0,\infty)$.
\end{definition}

CBF is a powerful tool used to handle control invariance of a set \cite{ames2019control}.
Suppose the set $C$ is given as a super zero level set of a $\mathcal{C}^1$ function $h:\mathbb{R}^n \rightarrow \mathbb{R}$, such that
\begin{equation} \label{eq:c_regularity}
\begin{aligned}
    C &= \{x\in \mathbb{R}^n : h(x) \geq 0\}, \\
    \partial C &= \{x\in \mathbb{R}^n : h(x) = 0\}, \\
    \mathrm{Int\;} C &= \{x\in \mathbb{R}^n : h(x) > 0\}.
\end{aligned}
\end{equation}

\begin{definition}[Control Barrier Function (CBF)] \label{def:cbf}
A continuously differentiable function $h:\mathbb{R}^n \rightarrow \mathbb{R}$ is a CBF for the dynamical system \autoref{eq:dynamics} if there exists a class $\mathcal{K}$ function\footnote{A function $\alpha: [0,\infty) \rightarrow \mathbb{R}$ belongs to class $\mathcal{K}$ if it is continuous, strictly increasing, and $\alpha(0)=0$.} $\alpha:[0,\infty) \rightarrow \mathbb{R}$ such that
\begin{equation} \label{eq:cbf_def}
    \exists u \in U \;\mathrm{s.t.,\;} {L_f h(x) + L_g h(x) \cdot u} + \alpha(x) \geq 0
\end{equation}
for all $x\in C$, and $\partial_x h(x) \neq 0$ for all $x \in \partial C$.
\end{definition}
\autoref{def:cbf} is also known by the name \textit{zeroing} CBF (ZCBF) \cite{ames2016control}, in the sense that the value of $h$ drops to zero when it approaches the boundary of set $C$. It should be clearly noted that a CBF can be defined only when the class $\mathcal{K}$ function $\alpha$ is specified. 

Now, let
\begin{equation} \label{eq:cbf_input_constraint_set}
    U_\mathrm{cbf} (x) = \{u \in U : L_f h(x) + L_g h(x) \cdot u + \alpha(x) \geq 0\}.
\end{equation}
It can be easily seen that \autoref{def:cbf} ensures $U_\mathrm{cbf}(x)$ is nonempty for all $x \in C$. Any feedback controller $u(x,t)$ will render the set $C$ invariant if $u(x,t) \in C$ for all $t \in [0,\infty)$. The following theorem is a result from Nagumo's theorem \cite[Section 4.2]{blanchini2008set}.

\begin{theorem}[Invariance Guarantees thorough CBF] \label{thm:cbf}
Let $u(x,t)$ be a Lipschitz continuous feedback law such that $u(x, t) \in U_\mathrm{cbf}(x)$ for all $x \in C$ and $t \in [0, \infty)$.
If a state trajectory $\mathrm{x}:[0,\infty)\rightarrow \mathbb{R}^n$ solves $\partial_t{\mathrm{x}}(t) = f(\mathrm{x}(t)) + g(\mathrm{x}(t)) u(\mathrm{x}(t), t)$ and $\mathrm{x}(0) \in C$, then $\mathrm{x}(t) \in C$ for all $t \in [0,\infty)$.
\end{theorem}

One widely-used way of synthesizing a control law $u(x,t)$ that satisfies the CBF constraint is to formulate an optimization problem:
\begin{equation}
    u(t,x) = \arg \min_{u\in U} \; J(x,u,t),
\end{equation}
subject to
\begin{equation}
    L_f h(x) + L_g h(x) \cdot u + \alpha(h(x)) \geq 0,
\end{equation}
where $J(x,u,t)$ is a convex cost function quadratic with respect to $u$.

The most popular choice of the cost function is
\begin{equation}
	J(x,u,t) = \lVert u - u_\mathrm{ref}(t,x) \rVert^2,
\end{equation}
where $\lVert\cdot\rVert$ is a weighted two-norm, $u_\mathrm{ref}(\cdot, \cdot)$ is a given (possibly feedback) reference input signal.
Since CBF-QP just adds one inequality constraint to the input constraint \autoref{eq:input_set_assumption}, it can be solved at a relatively low computational cost.

\section{Parametrized CBF} \label{sec:pcbf}
\subsection{Definition}

Given the basic definitions and the necessary assumptions, let us begin the discussion by setting up the problem in more detail.
Let $A \subset \mathbb{R}^n$ be the set of allowable states. For example, for a mobile robot collision avoidance task, $A$ contains all robot states not overlapping with the obstacles.
While $A$ is not necessarily control invariant, not every state in $A$ is actually safe, and we want a preferably large control invariant set $C$ such that $C \subseteq A$. As mentioned in the introduction, directly searching for a CBF (and also a corresponding class $\mathcal{K}$ function) for this purpose often becomes computationally intractable. This is because it requires solving the variational inequality \autoref{eq:cbf_def} subject to the inclusion condition $C \subseteq A$, i.e., 
\begin{equation} \label{eq:set_inclusion}
    h(x) \geq 0 \quad \Longrightarrow \quad x \in A.
\end{equation}

On the other hand, constructing a simple CBF without considering the constraint $C \subseteq A$ is relatively simpler in many cases. Moreover, it is often easy to obtain a \textit{continuous spectrum} of CBFs, rather than a single one.
That is, we can often find a single scalar function $h:\mathbb{R}^n \times K_0 \rightarrow \mathbb{R}$ such that for any $k \in K_0 \subseteq \mathbb{R}^{n_k}$, $h(\cdot, k)$ is a valid CBF that satisfies \autoref{eq:cbf_def}. We call such $h$ a PCBF.
\begin{definition}[Parametrized CBF]\label{def:pcbf}
A function $h:\mathbb{R}^n \times \mathbb{R}^{n_k} \rightarrow \mathbb{R}$ is a PCBF if there exists a continuous function $\alpha:\mathbb{R}\times \mathbb{R}^{n_k} \rightarrow \mathbb{R}$ such that, $\alpha(\cdot, k)$ is a class $\mathcal{K}$ function that makes $h(\cdot, k)$ a valid CBF for every fixed $k\in K_0 \subseteq \mathbb{R}^{n_k}$.
\end{definition}

\begin{remark} \label{remark:pcbf_is_a_generalization}
    PCBF is a generalization of CBF: Consider the case where $K_0$ is a set with only one element.
\end{remark}

Since $h(\cdot, k)$ is a valid CBF for every $k\in K_0$,
\begin{equation}
    C(k) = \{x \in \mathbb{R}^n : h(x,k) \geq 0\}
\end{equation}
is a control invariant set. Further, since the union of control invariant sets is also control invariant \cite[Proposition 4.13]{blanchini2008set}, for any $K \subseteq K_0$,
\begin{equation} \label{eq:C_def_pcbf}
    C = \bigcup_{k \in K} C(k) \subseteq \mathbb{R}^n
\end{equation}
is also control invariant.
While the sets $C(k)$ are not necessarily safe (i.e., $C(k) \subseteq A$), we can properly select the subset $K \subseteq K_0$ such that $C$ is not only control invariant but also $C \subseteq A$, without the need for directly addressing the set inclusion condition \autoref{eq:set_inclusion}. In that sense, $C(k)$ are \textit{building-block} control invariant sets whose union can form a large control invariant set which covers the entire workspace which is potentially large and complex.

Notice that finding a valid PCBF does not depend on the given environment $A$. Additionally, given a PCBF $h$, constructing a valid parameter constraint $K$ that matches the safety bound $A$ does not require explicit consideration of the dynamics. This enables decoupling of the barrier synthesis problem, as delineated in \autoref{prob:pcbf_design} and \autoref{prob:parameter_design}.

\subsection{Invariance-Preserving Control using PCBF} \label{sec:pcbfqp}

Now, we propose PCBF-QP, a QP-based computationally light safety filter that renders the PCBF invarinat set $C$ (of \autoref{eq:C_def_pcbf}) invariant, thereby constraining the state within $A$.
This is equivalent to driving the state trajectory $\mathrm{x}(t)$ in a way that there exists at least one choice of parameter $k \in K$ (which could indeed be time-varying) such that $\mathrm{x}(t) \in C(k)$.

A simple but naive approach for this is to pick a parameter $k \in K$ which makes $h(x,k) \geq 0$, along with the control input $u$ that satisfies the CBF constraint. That is, similarly to \cite{jang2024safe}, select $\mathrm{u}(t)$ and $\mathrm{k}(t)$ that solve
\begin{equation} \label{eq:naive_pcbf}
\begin{aligned}
    \min_{u, k} \; & J(x,k,u,t) \\
    \mathrm{s.t.}\; & L_f h(x,k) + L_g h(x,k) \cdot u + \alpha(h(x,k), k) \geq 0 \\
    & h(x,k) \geq 0 \\
    & \rho_i(k) \geq 0, \quad \forall i \in I \\
    & u \in U.
\end{aligned}
\end{equation}
with $x=\mathrm{x}(t)$, where $J$ is an appropriate cost function.
The problem in this form is that
\autoref{eq:naive_pcbf} is generally a nonlinear and nonconvex problem whose optimization typically requires heavy computation, making it inappropriate for real-time feedback synthesis.
Moreover, it is very likely that the solution is discontinuous with respect to $\mathrm{x}(t)$, which may result in a severe chattering phenomenon.

Thus, instead of directly optimizing over the parameter space, we will require the parameter $k$ to \textit{evolve continuously} with respect to time by \textit{controlling} it with through its time derivative. Consider the following augmented system
\begin{equation} \label{eq:augmented_dynamics}
\begin{aligned}
        \dot{\mathrm{x}}(t) &= f(\mathrm{x}(t)) + g(\mathrm{x}(t)) \mathrm{u}(t), \\
        \dot{\mathrm{k}}(t) &= \mathrm{v}(t),
\end{aligned}
\end{equation}
where $(\mathrm{x}(t),\mathrm{k}(t)) \in \mathbb{R}^n \times \mathbb{R}^{n_k}$ is the augmented state, $(\mathrm{u}(t),\mathrm{v}(t)) \in \bar{U} = U \times \mathbb{R}^{n_k}$ is the augmented input. Now, the parameter $k$ is no longer an optimization variable but the controller's internal state that has to be controlled by the virtual input $v \in \mathbb{R}^{n_k}$.

At this point, we introduce an additional (yet not restrictive) assumption that $K$ can be represented using differentiable inequality constraints, i.e., $K=\{k\in \mathbb{R}^{n_k}: \rho_i (k) \geq 0,\; \forall i\in I\}$, where $\rho_i$ are continuously differentiable functions, $I$ is an index set, with the regularity condition $\partial_k \rho_i(k) \neq 0$ if $\rho_i(k) = 0$.
With respect to the augmented system \autoref{eq:augmented_dynamics}, consider the disjoint union of $C(k)$, i.e.,
\begin{equation}
\begin{aligned}
    \bar{C} &= \bigsqcup_{k \in K} C(k) = \{(x,k) : k\in K,\; h(x,k) \geq 0\} \\
    &= \{(x,k) : \rho_{i}(k) \geq 0, h(x,k) \geq 0, \forall i \in I\}.
\end{aligned}
\end{equation}
Since $C(k)$ is a control invariant set with respect to the original system for every $k\in K$, $\bar{C}$ is control invariant with respect to the augmented dynamics: Consider $v=0$.
We will construct a constraint on the augmented input to render $\bar{C}$ invariant with respect to the augmented system. Since the projection of $\bar{C}$ onto the original state space $\mathbb{R}^n$ is $C$, invariance of $\bar{C}$ directly relates to invariance of $C$.

Nagumo's theorem \cite[Section 4.2]{blanchini2008set} says $\bar{C}$ is made invariant if
\begin{equation}
\begin{aligned}
    \rho_i(\mathrm{k}(t)) = 0 \;&\Rightarrow\; {\frac{d}{dt}}\rho_i(\mathrm{k}(t)) \geq 0 \quad \forall i \in I, \\
    h(\mathrm{x}(t), \mathrm{k}(t)) = 0 \;&\Rightarrow\; {\frac{d}{dt}}h(\mathrm{x}(t), \mathrm{k}(t)) \geq 0.
\end{aligned}
\end{equation}
Following the motivation of barrier function approaches including CBFs, we \textit{smoothen} this requirement by introducing the inequality constraint 
\begin{equation} \label{eq:kdot_constraint}
    \dot{\rho}_i(k,v) + \beta_i (\rho_i(k)) = \partial_k \rho_i (k) \cdot v + \beta_i (\rho_i(k)) \geq 0,
\end{equation}
for all $i\in I$.
Here, $\beta_i$ is a class $\mathcal{K}$ function which can be freely tuned to balance numerical stability and smoothness.
Similarly, for $h$ to be kept nonnegative, we require
\begin{equation} \label{eq:u_constraint}
\begin{aligned}
    &\frac{d}{dt}h(x,k) + \alpha(h(x,k), k) \\
    &= L_f h(x,k) + L_g h(x,k) \cdot u + \partial_k h(x, k) \cdot v + \alpha(h(x,k), k) \\
    &\geq 0.
\end{aligned}
\end{equation}
Notice the difference between \autoref{eq:u_constraint} and the original CBF constraint (the first constraint of \autoref{eq:naive_pcbf}). The added term $\partial_k h(x,k) \cdot v$ allows the controller to select the control $u$ from a wider range, compared to \autoref{eq:naive_pcbf}.

Combining the two requirements \autoref{eq:kdot_constraint} and \autoref{eq:u_constraint}, we can consider the following feasible set on the augmented input space:
\begin{equation} \label{eq:pcbf_input_set}
\begin{aligned}
    &\bar{U}_\mathrm{pcbf}(x,k) = \\ 
    &\left\{(u,v) \in \bar{U} :
    \begin{aligned}
            & L_f h + L_g h\cdot u + \partial_k h \cdot v + \alpha(h, k) \geq 0, \\
            & \partial_k \rho_i\cdot v + \beta_i(\rho_i) \geq 0, \; \forall i \in I
    \end{aligned}\right\},
\end{aligned}
\end{equation}
where (and hereafter when needed) the arguments of $h$ and $\rho_i$ are omitted for brevity.
The set $\bar{U}_\mathrm{pcbf}(x,k)$ is nonempty for all $(x,k) \in \bar{C}$. Let $v = 0$ and pick $u$ from the CBF input constraint set \autoref{eq:cbf_input_constraint_set} built upon the CBF $h(\cdot, k)$.

\begin{remark} \label{remark:time_varying}
    PCBF can also handle time-varying parameter constraints \textit{relaxing} with respect to time, for example, an exploration task.
    This means that $K$ is a set-valued function of time $K(t)$ and $K(t_1)\subseteq K(t_2)$ if $t_1 < t_2$.
    A sufficient yet not conservative condition for this is the existence of a class $\mathcal{K}$ function $\beta_i$ for every $i\in I$ such that $\partial_t \rho_i(k,t) + \beta_i(\rho_i(k,t)) \geq 0$.
    Then, replacing \autoref{eq:kdot_constraint} with $\partial_t \rho_i + \partial_k \rho_i \cdot v + \beta_i(\rho_i) \geq 0$ does not break invariance, since $v=0$ is still a feasible solution.
\end{remark}

\subsection{PCBF-based QPs for Invariance Guarantees}

Consider the following optimization-based controller, which we call PCBF-based QP (PCBF-QP):
Given the augmented dynamical system \autoref{eq:augmented_dynamics} and PCBF $h$, solve
\begin{equation} \label{eq:pcbf_qp}
\begin{aligned}
(\mathrm{u}(t), \mathrm{v}(t)) = \arg \min_{(u,v) \in \bar{U}} & J(\mathrm{x}(t),\mathrm{k}(t),u,v,t) \\
\mathrm{s.t.}\;\; & (u,v) \in \bar{U}_\mathrm{pcbf}(\mathrm{x}(t),\mathrm{k}(t)),
\end{aligned}
\end{equation}
where $J(\cdots)$ is a cost function that is strictly convex quadratic with respect to $(u, v)$. This QP always has a unique global minimizer for any $(\mathrm{x}(t),\mathrm{k}(t)) \in \bar{C}$, since $\bar{U}_\mathrm{pcbf}(\mathrm{x}(t),\mathrm{k}(t))$ is nonempty (as mentioned above) and $J$ is strictly convex with respect to the decision variables.
A decent choice of the cost function $J$ that works fine for many cases is to let $J$ take the CBF-QP-like form:
\begin{equation} \label{eq:cbf_qp_j}
	J(x,k,u,v,t) = \lVert u - u_\mathrm{ref}(x,t) \rVert_W^2 + \mu \lVert v \rVert_V^2,
\end{equation}
where $\lVert\cdot\rVert_W$ and $\lVert\cdot\rVert_V$ are weighted two-norms, $u_\mathrm{ref}$ is the reference input signal, $\mu > 0$ is a tunable and usually very small parameter which provides strict convexity and enhances numerical stability of the controller.

\begin{remark}[PCBF-QP with Infinitely Many Parameter Constraints] \label{remark:infinite_constraints}
It is interesting that the optimization problem \autoref{eq:pcbf_qp} remains convex and the same feasibility and invariance properties hold even with infinite number of parameter constraints, i.e., $I$ has infinite elements.
This suggests the potential of PCBF-QP to encompass a broader range of applications beyond simple QPs.
In \autoref{sec:semidefinite_example}, we show a case where infinitely many parameter constraints are employed and the resulting PCBF-QP is formulated into a semidefinite program (SDP) given a mild additional assumption on the class $\mathcal{K}$ function $\beta_{(\cdot)}$.
\end{remark}

\section{PCBF with High Relative Degree} \label{sec:hopcbf}

\begin{definition}[{Relative Degree \cite[Definition 13.2]{khalil2002nonlinear}}]
The output $h:\mathbb{R}^n \rightarrow \mathbb{R}$ of the system \autoref{eq:dynamics} has relative degree $r$ ($1 \leq r \leq n$) if $L_g L_f^{j} h(x) = 0$ for all $x \in \mathbb{R}^n$ and $j \in \{0, \cdots, r - 2\}$, and $L_g L_f^{r - 1} h(x) \neq 0$ almost everywhere (a.e.) on $\mathbb{R}^n$.
\end{definition}

Recently, HOCBFs \cite{xiao2021high} showed to be powerful when dealing with safety constraints of high (greater than one) relative degree. 
In this section, we extend the concept of PCBF to continuously parametrized safety constraints with high relative degree, which we call HOPCBF. Then, a control strategy to generate invariance-guaranteeing input at low computational cost based on QP, namely HOPCBF-QP, is derived.

\subsection{Review of HOCBF} \label{sec:hocbf_review}

We first begin with a brief overview of the motivation and theoretical details of HOCBF \cite{xiao2021high}.
Let $h:\mathbb{R}^n \rightarrow \mathbb{R}$ be a $\mathcal{C}^r$ output for the system \autoref{eq:dynamics} with relative degree $r$.
With this, we want to drive the system in a way that $h(\mathrm{x}(t)) \geq 0$ is satisfied throughout the interval $t \in [0,\infty)$.
The vector field $f$ appearing in the dynamics \autoref{eq:dynamics} is assumed to be at least $\mathcal{C}^{r-1}$, with its $(r-1)$-th order derivative being Lipschitz. 

Define a sequence of functions $\psi_{(\cdot)}:\mathbb{R}^n \rightarrow \mathbb{R}$ as follows:
\begin{equation} \label{eq:psi_def}
	\begin{aligned}
		\psi_0(x) &= h(x) \\
		\psi_j(x) &= \dot{\psi}_{j - 1}(x) + \alpha_j (\psi_{j - 1}(x)), \; \forall j \in \{1,\cdots,r - 1\}
	\end{aligned}
\end{equation}
where $\alpha_j(\cdot)$, $j\in \{1,\cdots,r - 1\}$ are class $\mathcal{K}$ functions.
For the sake of well-definedness of $\psi_j$, $\alpha_j$ is assumed to be at least $\mathcal{C}^{r - j}$.
Note that the relative degree of $\psi_j$ is at least $r - j$ and thus the term $\dot{\psi}_{j - 1}$ in the second line can be written as a function of $x$ only: $\dot{\psi}_{j-1}(x) = L_f \psi_{j-1}(x)$.
The key idea of HOCBF is that
\begin{equation} \label{eq:hocbf_chain}
    \psi_{j}(x) = \dot{\psi}_{j - 1}(x) + \alpha_j (\psi_{j - 1}(x)) \geq 0
\end{equation}
ensures $\psi_{j - 1}(x)$ value is kept nonnegative if it starts from a nonnegative initial value. If there exists a class $\mathcal{K}$ function $\alpha_r$ such that
\begin{equation} \label{eq:psi_r}
\begin{aligned}
&\max_{u \in U} \;\dot{\psi}_{r - 1}(x,u) + \alpha_r(\psi_{r - 1}(x)) \\
&{} = \max_{u \in U} \; L_f \psi_{r - 1}(x) + L_g \psi_{r - 1}(x) \cdot u + \alpha_r(\psi_{r - 1}(x)) \geq 0
\end{aligned}
\end{equation}
holds for all $x$ given $\psi_{(\cdot)}(x) \geq 0$,
then it will initiate a chain of nonnegativity certificates and eventually render the set 
\begin{equation} \label{eq:C_hocbf_def}
    C = \bigcap_{j = 1}^{r} C_j
\end{equation}
invariant, where for each $j \in \{1,\cdots,r\}$ the set $C_j$ is defined as $C_j = \{x \in \mathbb{R}^n : \psi_{j - 1} (x) \geq 0\}$.

\begin{definition}[High-Order CBF \cite{xiao2021high}] \label{def:hocbf}
    A function $h:\mathbb{R}^n \rightarrow \mathbb{R}$ having relative degree $r$ with respect to the system \autoref{eq:dynamics} is a HOCBF, if there exist class $\mathcal{K}$ functions $\alpha_j \in \mathcal{C}^{r - j}$, $j \in \{1,\cdots,r\}$ such that there exists an input $u \in U$ (depending on $x$) satisfying
    \begin{equation} \label{eq:hocbf_def}
        L_f \psi_{r - 1}(x) + L_g \psi_{r - 1}(x)\cdot u + \alpha_r(\psi_{r - 1}(x)) \geq 0,
    \end{equation}
    for all $x \in C$, where the functions $\psi_{(\cdot)}$ are defined recursively as $\psi_0(x) = h(x)$, $\psi_{j}(x) = L_f \psi_{j-1}(x) + \alpha_j(\psi_{j-1}(x))$ for $j \in \{1,\cdots,r-1\}$, and $C = \{x:\psi_{j-1}(x) \geq 0,\;\forall j \in \{1,\cdots,r\}\}$.
\end{definition}
\begin{remark}
    CBF is the special case $r=1$ of HOCBF.
\end{remark}

Now, let $U_\mathrm{hocbf} (x) = \{u \in U : \text{\autoref{eq:hocbf_def} holds.}\}$.
It can be easily seen that \autoref{def:hocbf} ensures nonemptyness of $U_\mathrm{hocbf}(x)$ for all $x \in C$. Any locally Lipschitz feedback controller $u(x,t)$ will render the set $C$ invariant if $u(x,t) \in U_\mathrm{hocbf}(x)$ for all $x \in C$ and $t \in [0,\infty)$. 
Leveraging this, one can generalize \cite{ames2016control} to high order:
\begin{equation}
\begin{aligned}
    u(t,x) = \arg \min_{u \in \mathbb{R}^m} \; & J(x, u, t) \\
    \mathrm{s.t.} \; & u \in U_\mathrm{hocbf}(x).
\end{aligned}
\end{equation}

\subsection{High-Order PCBF}

Now, we introduce the high-relative-degree version of PCBF and PCBF-QP similar to HOCBF, namely high-order PCBF (HOPCBF) and HOPCBF-QP. We begin with the follwing definition.

\begin{definition}[High-Order PCBF] \label{def:hopcbf}
    A function $h:\mathbb{R}^n\times \mathbb{R}^{n_k}\rightarrow \mathbb{R}$ is a HOPCBF of relative degree $r$ if there exist functions $\alpha_j\;(\in \mathcal{C}^{r-j}):\mathbb{R}\times \mathbb{R}^{n_k} \rightarrow \mathbb{R}$ ($j \in \{1,\cdots,r-1\}$) such that $\alpha_j(\cdot, k)$ are class $\mathcal{K}$ functions that make $h(\cdot, k)$ a valid HOCBF of relative degree $r$ for every fixed $k \in K$.
\end{definition}

With this definition, one might attempt to obtain the function sequence \autoref{eq:psi_def}. 
Unfortunately, considering the augmented system \autoref{eq:augmented_dynamics} with nonzero parameter change rate $\dot{k} \neq 0$, the HOPCBF \autoref{def:hopcbf} does not in general give well-defined function sequence $\psi_{(\cdot)}$ like the HOCBF's case, since there no longer is a guarantee that the time derivatives of $\psi_{(\cdot)}$ exist. This is because the parameter trajectory satisfying \autoref{eq:kdot_constraint} might not be $r$ times differentiable.
This is due to the mismatch in the relative degrees of the HOPCBF with respect to $\dot{k}$ (which is one) and to $u$ (which is $r$ by assumption).
Instead, we define another sequence of functions $\phi_{j-1}(x,k)$ ($j \in \{1,\cdots,r\}$) as
\begin{equation}
\begin{aligned}
    \phi_{0}(x,k) &= h(x,k) \\
    \phi_{j}(x,k) &= L_f \phi_{j - 1} (x,k) + \alpha_j(\phi_{j - 1}(x, k), k),
\end{aligned}
\end{equation}
and the sets
\begin{equation}
\begin{aligned}
    C_j(k) &= \{x\in \mathbb{R}^n : \phi_{j - 1} (x,k) \geq 0\}, \; j \in \{1, \cdots, r\}, \\
    C(k) &= \bigcap_{j=1}^r {C_j (k)}.
\end{aligned}
\end{equation}
The definitions for $\phi_{(\cdot)}$ are very similar to $\psi_{(\cdot)}$ from \autoref{eq:psi_def}, and they are actually identical given zero parameter speed (i.e., $\dot{k}=0$).
We set the control objective as to drive the system's state $x$ and the parameter $k$ such that $x \in C(k)$ and $k \in K$.
That is, we want to constrain the augmented system \autoref{eq:augmented_dynamics} within the set 
\begin{equation}
\begin{aligned}
    \bar{C} &= \bigsqcup_{k \in K} C(k) \\
    &= \left\{(x,k) : \begin{aligned}
    \rho_i(k) \geq 0 \quad &\forall i \in I, \\
    \phi_{j-1}(x,k) \geq 0 \quad &\forall j \in \{1,\cdots,r\}
    \end{aligned}\right\}
\end{aligned}
\end{equation}
by keeping $\phi_{j - 1}(\cdot)$ and $\rho_i(\cdot)$ values nonnegative.
Since $C(k)$ is a control invariant set with respect to the original system for every $k\in K$, $\bar{C}$ is control invariant with respect to the augmented dynamics: Consider $v=0$ under which $k$ would remain unchanged and $x$ will always be inside $C(k)$.

Similarly to PCBF with relative degree 1, we use Nagumo's theorem to construct a condition that renders $\bar{C}$ invariant:
\begin{equation}
\begin{aligned}
    \rho_i(\mathrm{k}(t)) = 0 \;&\Rightarrow\; {\frac{d}{dt}}\rho_i(\mathrm{k}(t)) \geq 0, \\
    \phi_{j-1}(\mathrm{x}(t), \mathrm{k}(t)) = 0 \;&\Rightarrow\; {\frac{d}{dt}}\phi_{j-1}(\mathrm{x}(t), \mathrm{k}(t)) \geq 0,
\end{aligned}
\end{equation}
for all $i\in I$ and $j\in \{1,\cdots,r\}$, which can be satisfied through the smoothened constraints on the augmented input space:
\begin{equation} \label{eq:hopcbf_kdot_constraint}
    \dot{\rho}_i(k,v) + \beta_i (\rho_i(k)) = \partial_k \rho_i (k) v + \beta_i (\rho_i(k)) \geq 0
\end{equation}
for all $i\in I$ which is the same to the relative-degree-one case,
\begin{equation} \label{eq:hopcbf_constraint_1}
\begin{aligned}
    & \dot{\phi}_{j - 1}(x,k,u,v) + \alpha_{j}(\phi_{j - 1}(x,k), k) \\
    &= L_f \phi_{j - 1}(x,k) + L_g \phi_{j - 1}(x,k) u + \partial_k \phi_{j - 1}(x,k) v \\
    & \quad {}  + \alpha_j(\phi_{j - 1}(x,k), k) \\
    &= \phi_{j}(x,k) + \partial_k \phi_{j - 1}(x,k) v \geq 0
\end{aligned}
\end{equation}
for all $j \in \{1,\cdots, r-1\}$, and
\begin{equation} \label{eq:hopcbf_constraint_2}
\begin{aligned}
    & \dot{\phi}_{r-1}(x,k,u,v) + \alpha_r (\phi_{r - 1}(x,k), k) \\
    &{}={} L_f \phi_{r - 1}(x,k) + L_g \phi_{r - 1}(x, k) u + \partial_k \phi_{r - 1}(x,k) v \\
    &\quad  {} + \alpha_r (\phi_{r - 1}(x,k), k) \geq 0.
\end{aligned}
\end{equation}
To obtain the last equality of \autoref{eq:hopcbf_constraint_1}, we used the fact that $\phi_j(x,k) = L_f \phi_{j -1}(x,k) + \alpha_j (\phi_{j - 1}(x,k))$, and the term $L_g \phi_{j - 1}(x,k)\cdot u$ reduces to zero since the relative degree of $\phi_{j - 1}(\cdot, k)$ with respect to the original dynamics \autoref{eq:dynamics} is at least $r - j + 1$, which is greater than $1$.

Combining \autoref{eq:hopcbf_kdot_constraint}, \autoref{eq:hopcbf_constraint_1}, and \autoref{eq:hopcbf_constraint_2}, we can consider the following feasible set on the augmented input space:
\begin{equation} \label{eq: hopcbf_constraint_set}
\begin{aligned}
    &\bar{U}_\mathrm{hopcbf}(x,k) = \\
    &\left\{(u,v) \in \bar{U} : \; \begin{aligned}
        &\text{\autoref{eq:hopcbf_constraint_1} $\forall j \in \{1,\cdots,r-1\}$, \autoref{eq:hopcbf_constraint_2},} \\
        &\partial_k \rho_i(k)\cdot v + \beta_i (\rho_i(k)) \geq 0, \; \forall i \in I
    \end{aligned}\right\},
\end{aligned}
\end{equation}
which is nonempty for any $(x,k) \in \bar{C}$.
This is because for any $u$ such that $L_f \phi_{r - 1}(x,k) + L_g \phi_{r - 1}(x,k) u + \alpha_r(\phi_{r-1}(x,k),k) \geq 0$, it can be easily seen that $(u,0) \in \bar{U}_\mathrm{hopcbf}(x,k)$. The existence of such $u$ is guaranteed for all $(x,k) \in \bar{C}$ since $h(\cdot, k)$ is a HOCBF.

\begin{remark}
HOPCBF does not have relative degree $r$ with respect to the augmented dynamics and is not an $r$-th order HOCBF for the augmented dynamics. This means we do not need to solve for high-order time derivatives of $\mathrm{k}(t)$ even with $r>1$. In return, we need $r$ inequality constraints on the augmented input to constrain the system within $\bar{C}$.
Another perspective of viewing this is that with respect to the augmented dynamics, we have constructed $r + |I|$ barrier-like functions, all with relative degree $1$, the intersection of whose super zero level sets defines $\bar{C}$, in a way that they are all compatible within $\bar{C}$, i.e., $\bar{U}_{\mathrm{hopcbf}}(x,k)$ is nonempty for all $(x,k) \in \bar{C}$. This compatibility is not general for naive composition of multiple (HO)CBFs, as there might not exist a control input that satisfies all the input constraints: The intersection of multiple control invariant sets is in general not invariant.
\end{remark}

\subsection{HOPCBF-QP}

Similar to PCBF-QP, a good way of synthesizing a feedback law that satisfies the HOPCBF constraints is to formulate an optimization problem
\begin{equation} \label{eq:pcbf_qp}
\begin{aligned}
(\mathrm{u}(t), \mathrm{v}(t)) = \arg \min_{(u,v) \in \bar{U}} & J(\mathrm{x}(t),\mathrm{k}(t),u,v,t) \\
\mathrm{s.t.}\;\; & (u,v) \in \bar{U}_\mathrm{hopcbf}(\mathrm{x}(t),\mathrm{k}(t)),
\end{aligned}
\end{equation}
which searches for the augmented input that minimizes the cost functional $J$. Again, this optimization could be solved at a relatively low computational cost if it becomes a QP, i.e., $J$ is strictly convex quadratic with respect to the decision variables $u$ and $v$, for example, \autoref{eq:cbf_qp_j}.

\begin{remark}
    In the same sense to \autoref{remark:pcbf_is_a_generalization}, HOPCBF is a generalization of HOCBF.
    It is also a high-relative-degree generalization of PCBF: The $r=1$ case will recover PCBF.
\end{remark}

\section{PCBF and Parameter Set Design} \label{sec:remarks}

Finding a general framework to construct a PCBF and the parameter constraints is not straightforward remains an open problem. 
However, in this section, we introduce some design techniques that apply to a class of practical dynamical systems.
Note that the discussions in this section applies to not only PCBFs but also HOPCBFs with $r \geq 2$.

\subsection{Symmetry-Induced PCBFs} \label{remark:pcbf_construction}

\begin{definition}[Continuous Symmetry of Dynamics] \label{def:continuous_symmetry}
The dynamics \autoref{eq:dynamics} is said to have continuous symmetry if there exists a Lie group $G$ acting on the state space $\mathbb{R}^n$ such that for any $q \in G$, if the state and input trajectory pair $(\mathrm{x},\mathrm{u})$ solves the ODE \autoref{eq:dynamics}, then also does the pair $(q\circ \mathrm{x}, \mathrm{u})$. Here, $q\in G$ as a function ($q:\mathbb{R}^n\rightarrow \mathbb{R}^n$) denotes the Lie group action.
\end{definition}

A system model found in real-world often (and almost always for a mobile robot) exhibits a continuous symmetry. 
This means that the dynamics is invariant under a continuous spectrum of coordinate changes. 
For example, the kinematics and dynamics of a planar mobile robot can be written in the same form regardless of which $SE(2)$ coordinate we choose.
As delineated in \autoref{def:continuous_symmetry}, continuous symmetries can be mathematically understood using Lie group actions. Continuous symmetry provides a simple yet powerful way of constructing (HO)PCBFs, which we call symmetry-induced PCBFs.

\begin{theorem}[Symmetry-Induced PCBF] \label{thm:symmetry_pcbf}
    Let $\hat{h}:\mathbb{R}^n \times \hat{K}_0 \rightarrow \mathbb{R}$ be a (HO)PCBF (with relative degree $r$, $\hat{K}_0 \subseteq \mathbb{R}^{\hat{n}_k}$) for a system with continuous symmetry with the corresponding Lie group of symmetry being $G$. Then, $h(x,k) = \hat{h}(q^{-1}(x), \hat{k})$ is a (HO)PCBF with the new parameter $k = (q, \hat{k}) \in K_0 = G \times \hat{K}_0$.
\end{theorem}
\begin{proof}
We will prove the equivalent statement: If $h:\mathbb{R}^n \rightarrow \mathbb{R}$ is a HOCBF, then $h' = h\circ q^{-1}$ also is.

Let $(\mathrm{x}, \mathrm{u}):[0,T) \rightarrow \mathbb{R}^n \times U$ be a dynamically feasible state-input trajectory pair where $T$ is a positive time horizon, i.e., $\mathrm{x}$ is the unique solution to the ODE $\dot{\mathrm{x}}(t) = f(\mathrm{x}(t)) + g(\mathrm{x}(t))\mathrm{u}(t)$ given initial condition $\mathrm{x}(0)$. Given the continuous symmetry and the same $\mathrm{u}$, if $\mathrm{z}:[0,T)$ solves the initial value problem $\dot{\mathrm{z}}(t) = f(\mathrm{z}(t)) + g(\mathrm{z}(t))\mathrm{u}(t)$ and $\mathrm{z}(0) = q(\mathrm{x}(0))$, then for all $t \in [0,T)$, $\mathrm{z}(t) = q(\mathrm{x}(t))$.

Thus, for any testing function $\beta:\mathbb{R}^n \rightarrow \mathbb{R}$, $\beta(\mathrm{x}(t)) = (\beta\circ q^{-1} \circ q \circ \mathrm{x})(t) = (\beta\circ q^{-1}) (\mathrm{z}(t))$. This means that under the same input signal $\mathrm{u}(t)$, $\left(\frac{d}{dt}\right)^j \beta(\mathrm{x}(t)) = \left(\frac{d}{dt}\right)^j (\beta\circ q^{-1}) (\mathrm{z}(t))$ up to any order $j \in \{0,1,\cdots\}$ as long as they exist. Letting $\beta$ be the $\phi_{(\cdot)}$ functions concludes the proof.
\end{proof}

Symmetry-induced PCBFs are also powerful in terms of parameter constraint construction, since the shape of the invariant sets $C(k)$ remain unchanged under the orbit of the Lie group action.
\begin{theorem} \label{thm:sym_pcbf_C}
    Suppose $h$ is a symmetry-induced PCBF with the Lie group of symmetry and the parameter space being $G$ and $G \times \hat{K}_0$, respectively. Let $C_e(\hat{k}) = C((1_G, \hat{k}))$, where $1_G$ is the identity element of $G$. Then, $C(k) = q(C_e(\hat{k})) = \{q(x) : x \in C_e(\hat{k})\}$ ($k = (q, \hat{k})$).
\end{theorem}
\begin{proof}
    The result follows directly from the definition of symmetry-induced PCBF $h(x,k) = \hat{h}(q^{-1}(x), \hat{k})$.
\end{proof}
\noindent \autoref{thm:sym_pcbf_C} tells that $C(k)$-s are \textit{transformed copies} of $C_e(\hat{k})$, which reduces the search space for constructing the parameter constraint. For example, if $A$ is expressed as
\begin{equation}
    A = \{x\in \mathbb{R}^n : l_i(x) \geq 0,\;\forall i \in I\},
\end{equation}
then any $\rho_i$ such that
\begin{equation}
    \rho_i(k) \leq \min_{x \in C_e(\hat{k})} (l_i \circ q) (x)
\end{equation}
would make $C \subseteq A$. Notice that the search space on the right hand side is reduced to $C_e(\hat{k})$.

\subsection{PCBF Construction using Stabilizing Control} \label{sec:lyapunov_pcbf}

Let $x_0 \in \mathbb{R}^n$ be a point in state space to which the system is stabilizable. That is, there exists a Lyapunov function $V(\cdot)$ such that $V(x) \geq 0$ for all $x$, $V(x_0)=0$ if and only if $x=x_0$, and
\begin{equation}
    \min_{u \in U} L_f V(x) + L_g V(x) u \leq 0.
\end{equation}
Then, for any $b \geq 0$, $b - V(x)$ is a CBF \cite{jang2024safe}, and therefore $h(x,b) = b-V(x)$ is a PCBF with parameter $b \in K_0 = \{b\in \mathbb{R} : b\geq 0\}$.
In many cases, finding a valid Lyapunov function is easier than searching directly for a (HO)CBF, not only because Lyapunov functions are sometimes handcraftable, but also because there are readily a handful of existing nonlinear control methodologies designed specifically for stabilization of nonlinear systems.
These include but not are limited to backstepping control, sliding mode control, neural Lyapunov control \cite{chang2019neural}, all coming with valid Lyapunov functions. With appropriate setting, Hamilton-Jacobi reachability \cite{bansal2017hamilton, bansal2021deepreach} method can also be used to find a Lyapunov function.

Combined with the symmetry-based technique of the previous subsection, this Lyapunov-based method serves as an especially powerful tool in tasks such as collision avoidance of mobile robots.
If the system exhibits a continuous symmetry, the system being stabilizable to $x_0 \in \mathbb{R}^n$ automatically implies it is also stabilizable to $q(x_0)$, for any $q$ from the group of symmetry $G$, and as such, $h(x,k)=b-V(q^{-1}(x))$ is a valid PCBF with the parameter $k$ being $k=(b \geq 0,q \in G)$.
In other words, the PCBF framework allows building a safety filter for complex environments using only a single stabilizing controller.

\subsection{Parameter Augmentation using Auxiliary Variables} \label{sec:slack}

Another advantage offered by the PCBF framework is that it allows augmenting the parameter space using \textit{auxiliary} variables.
That is, given a PCBF $h$, the user can build a new PCBF $\bar{h}(x,\bar{k}) = h(x,k)$ with the augmented parameter $\bar{k} = (k,\eta) \in \bar{K} = K_0 \times E$, where $\eta \in E \subseteq \mathbb{R}^{n_\eta}$ is the auxiliary variable.
One can adopt many constraint engineering techniques from mathematical optimization \cite{boyd2004convex} to mitigate the burden of building the parameter constraints.
For example, if the shape of $A$ is complex to handle, one can define a simple-shaped (e.g., polygonal or ellipsoidal) set-valued function $D(\eta) \subseteq \mathbb{R}^{n}$, and specify the parameter constraints to ensure $D(\eta) \subseteq A$ and $C(k) \subseteq D(\eta)$.

\section{Case Study} \label{sec:case_study}

In this section, we present two simulation results that well exemplify practical applications of the proposed framework.
The first scenario addresses a mobile robot collision avoidance problem, where the design techniques introduced in \autoref{sec:remarks} are used.
In the second scenario, we demonstrate HOPCBF-QP with $r = 2$ using a simplified longitudinal dynamics model of a vehicle, where a time-varying parameter constraint (see \autoref{remark:time_varying}) is employed.
We also showcase \autoref{remark:infinite_constraints} through the third example using a constrained stabilization task of a linear system.

\subsection{Collision-Free Mobile Robot Navigation} \label{sec:mobile_robot}

\begin{figure*}
    \centering
    \includegraphics[page=1, width=\linewidth]{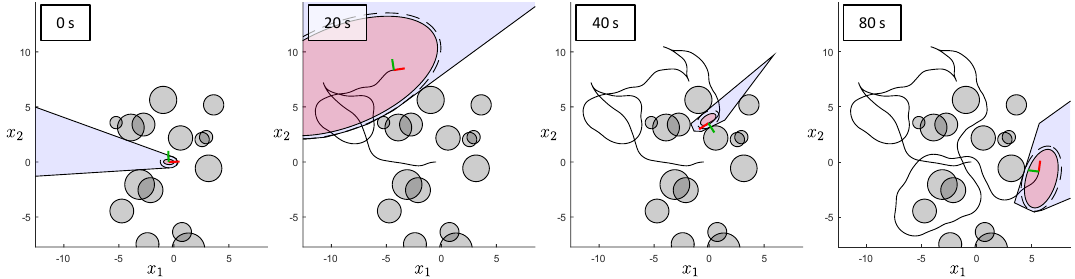}
    \caption{Four snapshots taken from the simulation on collision-free mobile robot navigation (\autoref{sec:mobile_robot}). The obstacle configuration (position and size) is randomly chosen, and the reference input $u_\mathrm{ref}$ is manually given by a human operator who is instructed to transmit aggressive inputs. In each subfigure, the red ellipse denotes $C(k)$, the dotted ellipse is $C(k)$ buffered by the robot's size, black solid line is the robot's trajectory on the $x_1$-$x_2$ plane, gray shaded regions are the obstacles, and the blue polygonal region is the collision-free space defined by the separating hyperplanes described by the auxiliary variable $\eta$. The boxes on the top left of each snapshot denotes the time the snapshot is taken.
    The attitude of the robot is depicted as red and green axes.
    PCBF-QP ensures the robot to stay away from any collision, regardless of the aggressiveness of the input.}
    \label{fig:mobile_robot_snapshots}
\end{figure*}

In this example, following the design techniques explained in \autoref{sec:remarks}, we will construct a PCBF for a wheeled ground rover navigating in obstacle-cluttered space obeying the following simplified bicycle-like dynamics:
\begin{equation} \label{eq:rover_dynamics}
    \dot{x} = \begin{bmatrix}
        \dot{x}_1 \\ \dot{x}_2 \\ \dot{x}_3 \\ \dot{x}_4
    \end{bmatrix} = f(x) + g(x)u = \begin{bmatrix}
        x_3 \cos x_4 \\ x_3 \sin x_4 \\ 0 \\ 0
    \end{bmatrix} + \begin{bmatrix}
        0 & 0 \\ 0 & 0 \\ 1 & 0 \\ 0 & x_3
    \end{bmatrix} \begin{bmatrix}
        u_1 \\ u_2
    \end{bmatrix},
\end{equation}
where the components of $x = [x_1, x_2, x_3, x_4]^\top \in \mathbb{R}^4$ denote the horizontal and vertical positions, forward velocity, and heading angle of the robot, respectively, which are controlled through acceleration ($u_1 \in \mathbb{R}$) and steering ($u_2 \in \mathbb{R}$) inputs. We assume that the inputs are bounded by a box constraint $u \in U = [-1, 1] \times [-1, 1]$.

The mission for this example is to track the reference input given by the user, while avoiding multiple circular shaped obstacles. The number of obstacles is $N$, and for each $i \in \{1,\cdots,N\}$, the $i$-th obstacle is located at position $(z_{i, 1}, z_{i, 2})$ on the $x_1$-$x_2$ plane and has radius $R_i>0$.
The robot is modeled as a circle on the $x_1$-$x_2$ plane, having radius $R>0$.

As the first step, we find that the dynamics \autoref{eq:rover_dynamics} is continuously symmetric under the Lie group action of $SE(2)$. This symmetry is very natural in that the dynamics of a ground robot \autoref{eq:rover_dynamics} can be written in the same form regardless of the choice of coordinate.
For a $q \in SE(2)$, the Lie group action $q(x)$ is defined as \textit{accordingly translating and rotating} the pose-related elements $(x_1, x_2, x_4)$ with the velocity $x_3$ remaining unchanged.
We also find that the robot is stabilizable to the origin by utilizing the following handcrated control Lyapunov function.
\begin{equation}
    V(x) = \sqrt{x_1^2 + 4x_2^2 + \epsilon^2} - \epsilon + \frac{1}{2}x_3^2 + 1 - \cos x_4
\end{equation}
This $V$ is a valid Lyapunov function for any $\epsilon > 0$ because
\begin{equation}
    u = \frac{1}{\sqrt{x_1^2 + 4x_2^2 + \epsilon^2}}\begin{bmatrix}
        -x_1 \cos x_4 - 2x_2 \sin x_4 \\
        -2x_2
    \end{bmatrix} \in U
\end{equation}
is a feasible input that makes the $V(x)$ value nonincreasing.
With the two observations given, we follow the steps delineated in \autoref{sec:remarks} to obtain a symmetry-induced PCBF
\begin{equation*}
    h(x,k) = b - V(q^{-1}(x)),\; k = (b,q) \in K_0 = [0,\infty) \times SE(2).
\end{equation*}

\begin{figure}
    \centering
    \includegraphics[page=1, width=\linewidth]{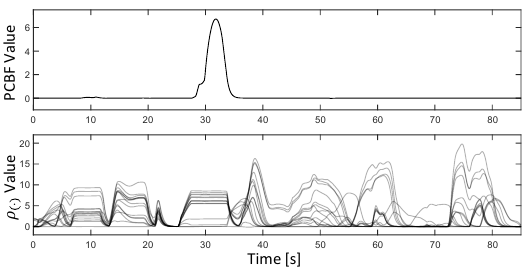}
    \caption{The values of PCBF $h$ and parameter constraint $\rho_{(\cdot)}$ in the robot navigation example, plotted as a function of time. It can be seen that PCBF-QP is capable of keeping the values nonnegative at all times.}
    \label{fig:mobile_robot_plots}
\end{figure}

Notice that the $x_1$-$x_2$ projection of the $b$-level set of $V(x)$, i.e., $C(k)$ with $q$ being the identity element, is always an ellipse with
the major semiaxis pointing to the positive $x_1$ direction.
Utilizing this and \autoref{thm:sym_pcbf_C}, and following \autoref{sec:slack}, we augment the parameter space using an $N$-dimensional auxiliary variable $\eta \in \mathbb{R}^N$.
This defines $N$ hyperplanes (i.e., lines) on the $x_1$-$x_2$ space, resulting in polygonal $D(\eta) \subseteq \mathbb{R}^2$. For each element of $\eta$, we employ a parameter constraint which constrains each hyperplane to strictly pass between each obstacle and the ellipse (buffered by the robot's size $R$), as shown in \autoref{fig:mobile_robot_snapshots}. This ensures $C(k)$ and the obstacles do not overlap, and thus $C \subseteq A$.
We omit the details of the derivation due to limited space and since it is a tedious series of basic hand-doable calculations.

Simulation experiment was conducted using $\epsilon = 0.01$, $R = 0.3$ and $N=15$ randomly placed obstacles of random sizes. Note that handcrafting a single CBF that covers this workspace is almost impossible. We used $\alpha(y,k) = 2y$ and $\beta_{(\cdot)}(y,k) = 2y$ for the class $\mathcal{K}$ functions.
The reference input $u_\mathrm{ref}$ is given through manual control by a human operator who is instructed to give aggressive inputs towards the obstacles, so the overall closed-loop system should rely on the PCBF to avoid any collision.
\autoref{fig:mobile_robot_snapshots} shows four snapshots taken from the simulation. Regardless of the aggressiveness of the manual reference input, the robot always stays within the set $C(k)$ which is placed collision-free due to the parameter constraints.
In \autoref{fig:mobile_robot_plots}, it can be seen that the values of PCBF $h$ and the parameter constraint functions $\rho_{(\cdot)}$ are kept nonnegative throughout the simulation.

\subsection{Inter-Vehicle Distance Maintenance} \label{sec:acc}

\begin{figure}
\centering
\includegraphics[page=1, width=79.7mm]{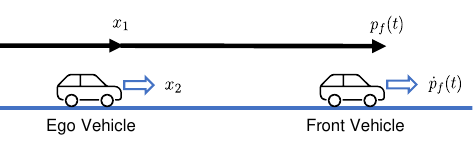}
\caption{The inter-vehicle distance maintenance scenario (\autoref{sec:acc}). The ego vehicle is required to maintain a safe distance of $\delta$ from the front vehicle. The position of the ego vehicle and the front car are $x_1$ and $p_f$, respectively. Their velocities are $x_2$ and $\dot{p}_f$.}
\label{fig:acc_car}
\end{figure}
\begin{figure*}
    \centering
    \includegraphics[page=1, width=\textwidth]{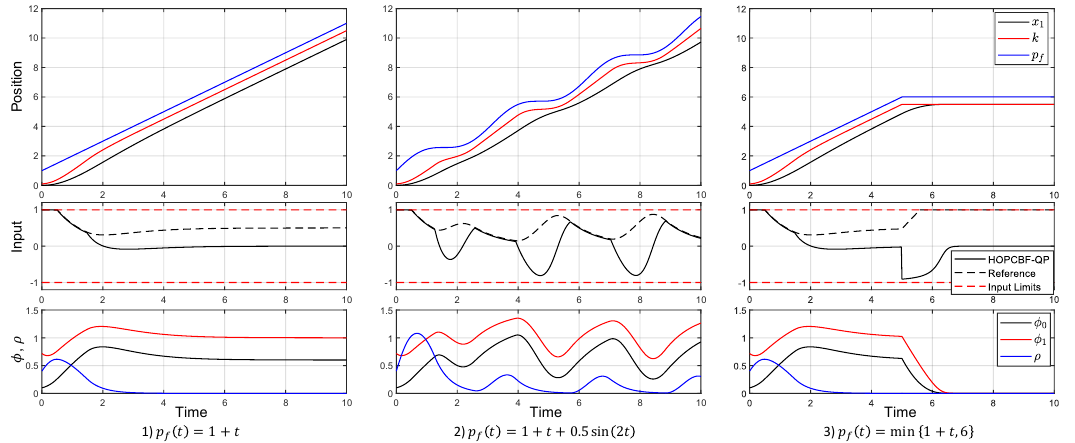}
    \caption{Simulation result for \autoref{sec:acc}. Regardless of the front vehicle behavior (as long as it does not reverse), HOPCBF-QP is capable of keeping the ego vehicle's position $x_1$ at least $\delta = 0.5$ apart from the front vehicle's position $p_f$. The values of $\phi_{(\cdot)}$ and $\rho$ are simultaneously kept nonnegative through HOPCBF-QP.}
    \label{fig:acc_result}
\end{figure*}

Consider the following simplified vehicle dynamics:
\begin{equation}
    \dot{x} = \begin{bmatrix}
        \dot{x}_1 \\ \dot{x}_2
    \end{bmatrix} = f(x) + g(x) u = \begin{bmatrix}
        0 & 1 \\ 0 & 0
    \end{bmatrix} \begin{bmatrix}
        x_1 \\ x_2
    \end{bmatrix} + \begin{bmatrix}
        0 \\ 1
    \end{bmatrix} u,
\end{equation}
where each component of the state $x=[x_1\;x_2]^\top \in \mathbb{R}^2$ denote the position and the velocity of the vehicle, $u\in [u_L, u_U] \subset \mathbb{R}$ denotes the acceleration input.
The two parameters $u_L < 0$ and $u_U > 0$ denote the control bounds.
As shown in \autoref{fig:acc_car}, the goal of the ego vehicle is to move forward at a prescribed speed $v_\mathrm{ref} > 0$ while maintaining a safe distance of $\delta > 0$ with the preceding vehicle at position $p_f(t)$ and velocity $\dot{p}_f(t)$. We assume that the front car never moves backwards, i.e., $\dot{p}_f(t)$ is nonnegative.

The HOPCBF $h$ is designed as follows:
\begin{equation} \label{eq:acc_hopcbf_cand}
    h(x,k) = k-x_1
\end{equation}
where $k \in \mathbb{R}$ is the parameter whose physical interpretation is the position on the road before which the vehicle is able to come to a complete stop.
With $\alpha_1(y,k) = \sqrt{ay + \epsilon^2} - \epsilon$ and $\alpha_2(y,k) = \gamma y$ ($y \geq 0$, $\epsilon > 0$, $\gamma > 0$), this $h$ satisfies the condition in \autoref{def:pcbf} if $a \geq -2/u_L>0$. With them, $\phi_{(\cdot)}$ are given as follows:
\begin{equation}
\begin{aligned}
    \phi_{0}(x,k) &= h(x,k) = k-x_1 \\
    \phi_{1}(x,k) &= -x_2 + \sqrt{a(k - x_1) + \epsilon^2} - \epsilon.
\end{aligned}
\end{equation}

To avoid collision with the front vehicle, we introduce one time-varying parameter constraint
\begin{equation}
    \rho(k, t) = p_f(t) - k - \delta.
\end{equation}
It is straightforward to check that $\rho(k,t) \geq 0$ if and only if $p_f(t) - x_1 \geq \delta$ for all $x = [x_1, x_2]^\top \in C(k)$, and $\rho(k,t)$ always increases with respect to time and thus the conditions in \autoref{remark:time_varying} holds with any class $\mathcal{K}$ function $\beta$.

In order to encourage the ego vehicle to move at a speed close to $v_\mathrm{ref}$, we make use of HOPCBF-QP with cost $J(x,k,u,v,t) = (\mathrm{sat}(u_\mathrm{ref}(x,t)) - u)^2 + \mu v^2$ where the reference input $u_\mathrm{ref}$ is given as a simple linear speed feedback $u_\mathrm{ref}(x,t) = L(v_\mathrm{ref} - x_2)$.
Here, $\mathrm{sat}(\cdot)$ is the saturation function that clips off the excessive input to fit the bound $u \in [u_L, u_U]$, and $\mu$ and $L$ are constant positive reals.

Simulation was conducted using $u_L = -1$, $u_U = 1$, $\delta = 0.5$, $\epsilon = 0.1$, $a = 2$, $\gamma = 2$, $\beta(y) = 2y$, $\mu = 0.01$ and $L = 1$. The ego vehicle starts at $x = 0$ and $k = 0.1$, and its reference speed is $v_\mathrm{ref} = 1.5$. For the leading vehicle behavior, we consider three different scenarios.
\begin{enumerate}
    \item $p_f(t) = 1 + t$: The front vehicle moves at a constant speed which is slower than $v_\mathrm{ref}$.
    \item $p_f(t) = 1 + t + 0.5 \sin (2t)$: The front vehicle repeatedly accelerates and decelerates.
    \item $p_f(t) = \min \{1 + t, 6\}$: The front vehicle first moves at a constant speed, and then suddenly stops at $t = 5$.
\end{enumerate}
The results for three scenarios can be found in \autoref{fig:acc_result}.
As shown in the plots, the ego vehicle successfully keeps the safe distance from the preceding vehicle and the input limits simultaneously through HOPCBF-QP.

\subsection{Constrained Stabilization of a Linear System} \label{sec:semidefinite_example}

In the second example, we consider the stabilization task of the following single-input single-output (SISO) linear time-invariant (LTI) system:
\begin{equation}
\begin{aligned}
	\dot{x} &= Ax + Bu \\
	y &= c^\top x,
\end{aligned}
\end{equation}
where $x \in \mathbb{R}^n$ is the state, $u, y \in \mathbb{R}$ are the single-dimensional input and output, $A$, $B$, $c$ are constant matrices and a column vector of appropriate sizes.
With the assumption that the given system is stabilizable, we want to stabilize it to the origin while keeping the output $y$ of the system within a prescribed bound near the origin with bounded input.
Since this system is LTI, without loss of generality, we can let the input and output constraints be $-1 \leq u \leq 1$ and $-1 \leq y \leq 1$, respectively.

To achieve this control objective, following \autoref{sec:lyapunov_pcbf} we set up the PCBF candidate function $h$ as
\begin{equation} \label{eq:lin_syst_pcbf}
h(x,k) = b - \frac{1}{2}x^\top Px,
\end{equation}
where $k = (b,L,P)$ is the parameter consisting of the Lyapunov function bound $b\in \mathbb{R}$, feedback gain $L \in \mathbb{R}^{1 \times n}$, and the Lyapunov function candidate $P \in \mathbb{S}^n$.
Similar to the role of $\eta_{(\cdot)}$ in the previous example, although the second component $L$ does not explicitly appear in \autoref{eq:lin_syst_pcbf}, it serves as a slack variable when formulating the parameter constraints.

For \autoref{eq:lin_syst_pcbf} to be a PCBF, similar to Lyapunov-based CBFs \cite{jang2024safe}, we require 
$V_P(x) = \frac{1}{2}x^\top P x$ be a Lyapunov function for the closed-loop system $\dot{x} = (A - BL)x$, i.e.,
\begin{equation}
    \dot{V}_P(x) = \frac{1}{2}x^\top \left(P(A-BL) + (A-BL)^\top P\right)x \leq 0
\end{equation}
for all $x$, and therefore we set the first parameter constraint as
\begin{equation} \label{eq:lti_rho1}
\rho_1 (k) (\in \mathbb{S}^n) = -P(A - BL) - (A - BL)^\top P \geq 0.
\end{equation}

Secondly, for the PCBF condition to be satisfied with bounded input, we require the feedback input $u = -Lx$ to meet the specified input bounds, i.e., $-1 \leq Lx \leq 1$ for all $x \in C(k) = \{x:\frac{1}{2}x^\top Px \leq b\}$.
This is equivalent to
\begin{equation} \label{eq:lti_rho2}
	\rho_2 (k) (\in \mathbb{S}^n) = P - 2bL^\top L \geq 0,
\end{equation}
which serves as the second parameter constraint.

Similarly, to satisfy the output constraint ($-1 \leq c^\top x \leq 1$ for all $x \in C(k)$), we let
\begin{equation} \label{eq:lti_rho3}
	\rho_3 (k) (\in \mathbb{S}^n) = P - 2bcc^\top \geq 0
\end{equation}
be the third parameter constraint.

Note that the parameter constraints \autoref{eq:lti_rho1}, \autoref{eq:lti_rho2}, and \autoref{eq:lti_rho3} are expressed as semidefinite constraints, rather than scalar inequalities.
A semidefinite constraint is equivalent to having a spectrum of infinite number of scalar inequality constraints that reads
\begin{equation}
\xi^\top \rho_{(\cdot)}(k) \xi \geq 0, \quad \forall \xi \in \mathbb{R}^{p},
\end{equation}
where $p$ is the appropriate dimension. Thus, the parameter velocity $v$ should satisfy
\begin{equation} \label{eq:lti_rhodot}
	\frac{d}{dt} \left(\xi^\top \rho_{i}(k) \xi\right) + \beta_{i} \left(\xi^\top \rho_{i}(k) \xi\right) \geq 0, \quad \forall \xi \in \mathbb{R}^p
\end{equation}
for all $i$.
To deal with this, we take linear class $\mathcal{K}$ functions $\beta_{(\cdot)}(z) = \gamma_{(\cdot)}\cdot z$ ($\gamma_{(\cdot)} > 0$, $z \in \mathbb{R}$), and then \autoref{eq:lti_rhodot} again takes the semidefinite form:
\begin{equation}
	\frac{d}{dt} \rho_{i}(k) + \gamma_{i} \cdot \rho_{i}(k) \geq 0,
\end{equation}
which yields semidefinite PCBF-QP.

To stabilize the system to the origin, we take a control Lyapunov function
\begin{equation}
    V(x) = \frac{1}{2}x^\top S x \quad (S \in \mathbb{S}_+^n)
\end{equation}
and then the following CLF-CBF-QP-style \cite{ames2016control} PCBF-QP, namely CLF-PCBF-QP, is employed.
\begin{equation}
\begin{aligned}
    \min_{u \in U,v,\delta} \quad & \lVert u \rVert^2 + \mu \lVert v \rVert^2 + \nu \delta^2 \\
    \mathrm{s.t.} \quad & L_f V(x) + L_g V(x)\cdot u + \alpha_{\mathrm{clf}}(V(x)) \leq \delta \\
    & L_f h(x,k) + L_g h(x,k)\cdot u + \partial_k h(x,k) \cdot v \\ 
    & \quad {} + \alpha(h(x,k)) \geq 0 \\
    & \frac{d}{dt}\rho_i(k) + \gamma_i \cdot \rho_i(k) \geq 0, \quad \forall i \in \{1, 2, 3\}
\end{aligned}
\end{equation}
Here, $\mu$ and $\nu$ are positive weights, $\alpha_\mathrm{clf}$ is a class $\mathcal{K}$ function, $\delta\in \mathbb{R}$ is an optimization variable that penalizes insufficient decaying speed of the CLF value without affecting the overall feasibility of the optimization.

\begin{figure}
    \centering
    \includegraphics[page=1, width=\linewidth]{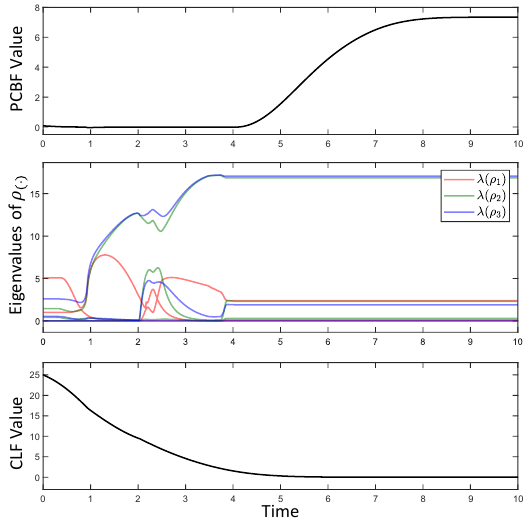}
    \caption{
    The values of PCBF $h$, parameter constraints $\rho_{(\cdot)}$, and CLF $V$ in the linear system control scenario (\autoref{sec:semidefinite_example}), plotted as a function of time. In the second plot, the eigenvalues of the parameter constraints are drawn. The parameter constraints are satisfied if and only if all the eigenvalues are nonnegative. It can be seen that the CLF value successfully decays towards zero while PCBF and $\rho_{(\cdot)}$ values being kept nonnegative through CLF-PCBF-QP.
    Note that $\rho_{(\cdot)}$ are guaranteed to have real eigenvalues since they are symmetric matrices.
    }
    \label{fig:linear_plot}
\end{figure}
\begin{figure}
    \centering
    \includegraphics[page=1]{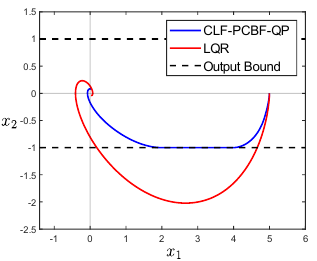}
    \caption{
    The trajectory portrait of the linear system control scenario, projected onto the $x_1$-$x_2$ plane. While the LQR-based controller results in violation of the output constraint, the proposed PCBF-based controller (CLF-PCBF-QP) successfully stabilizes the target system while satisfying the requirements.
    }
    \label{fig:linear_phase}
\end{figure}

Here, we consider the triple integrator model with system matrices
\begin{equation}
    A = \begin{bmatrix}
        0 & 1 & 0 \\
        0 & 0 & 1 \\
        0 & 0 & 0
    \end{bmatrix}, \; B = \begin{bmatrix}
        0 \\ 0 \\ 1
    \end{bmatrix}
\end{equation}
where the components of the three-dimensional state are written as $x = [x_1, x_2, x_3]^\top$.
The objective here is to bound its velocity component $x_2 = c^\top x \in [-1, 1]$ ($c = [0, 1, 0]^\top$).

The simulation results using $S = [2, 2, 1; 2, 3, 2; 1, 2, 2]$, $\mu = 0.1$, $\nu = 10$, $\alpha_\mathrm{clf}(y)=5y$, $\gamma_{(\cdot)} = 5$ starting from the initial state $\mathrm{x}(0) = [5, 0, 0]^\top$ are shown in \autoref{fig:linear_plot}. For the parameter constraints, their eigenvalues are depicted as they are semidefinite constraints. It can be seen that $\rho_{(\cdot)}$ values are kept semidefinite, and $h(x,k)$ nonnegative throughout the simulation.
\autoref{fig:linear_phase} shows the closed-loop trajectory projected onto the $x_1$-$x_2$ plane. We compare with linear quadratic regulator (LQR) controller which gives the same Lyapunov function $\frac{1}{2}x^\top Sx$. It shows how the velocity $x_2$ is well bounded through the deployment of CLF-PCBF-QP.

\section{Conclusion} \label{sec:conclusion}

In this work, we introduced the concept of PCBF, a differentiably parametrized spectrum of CBFs, along with PCBF-QP, a QP-based feedback controller that uses a PCBF. The concept was extended to high-order cases which is called HOPCBF.
Multiple parameter constraints can be addressed using a PCBF, allowing it to cover a relatively large and complex subset of the workspace using simple building-block control invariant sets.
We also introduced some design techniques that can be used for a class of systems to design a valid PCBF and the parameter constraints for invariance guarantees within a given safe region.
Simulation experiments were conducted to validate the proposed methodology.

While the proposed method was successful in building a control invariant set large enough to accomplish the task in the shown simulation environments, it remains an open question how we could systematically design a suitable PCBF for more general systems and tasks, warranting further research.
In addition, fusing with stochastic control methods to enable PCBFs to cover uncertain or stochastic dynamics models is another possible future work.

\bibliography{IEEEabrv, references}

\end{document}